# A Pipelined Memristive Neural Network Analog-to-Digital Converter

Loai Danial, Kanishka Sharma, and Shahar Kvatinsky
*Andrew and Erna Viterbi Faculty of Electrical Engineering,*
Technion - Israel Institute of Technology, Haifa 3200003, ISRAEL, Email: sloaidan@campus.technion.ac.il

*Abstract*— With the advent of high-speed, high-precision, and low-power mixed-signal systems, there is an ever-growing demand for accurate, fast, and energy-efficient analog-to-digital (ADCs) and digital-to-analog converters (DACs). Unfortunately, with the downscaling of CMOS technology, modern ADCs trade off speed, power and accuracy. Recently, memristive neuromorphic architectures of four-bit ADC/DAC have been proposed. Such converters can be trained in real-time using machine learning algorithms, to break through the speed-power-accuracy trade-off while optimizing the conversion performance for different applications. However, scaling such architectures above four bits is challenging. This paper proposes a scalable and modular neural network ADC architecture based on a pipeline of four-bit converters, preserving their inherent advantages in application reconfiguration, mismatch self-calibration, noise tolerance, and power optimization, while approaching higher resolution and throughput in penalty of latency. SPICE evaluation shows that an 8-bit pipelined ADC achieves 0.18 LSB INL, 0.20 LSB DNL, 7.6 ENOB, and 0.97 fJ/conv FOM. This work presents a significant step towards the realization of large-scale neuromorphic data converters.

## I. INTRODUCTION

High performance data converters are key components in modern mixed-signal systems, in advanced technology nodes, and emerging data-driven applications. However, the analog performance in the same process is dramatically degraded due to reduced signal-to-noise ratio (SNR), low intrinsic gain, device leakage, and device mismatch [1]. These deep-submicron effects exacerbate the intrinsic speed-power-accuracy tradeoff in the Analog-to-digital converters (ADCs), which has become a chronic bottleneck of modern system design [2]. Moreover, these effects are poorly handled with specific and time-consuming design techniques for special purpose applications, resulting in considerable overhead and severely degrading their performance [2].

For example, the pipelined architecture of ADCs achieves high resolution at a moderate-high speed, but conventional designs rely on proper component matching and require complex op-amps which are increasingly difficult to design and scale in state-of-the-art CMOS technologies [3], [4]. Additionally, their flash-type sub-ADCs require high power and area due to a large number of accurate comparators, pushing them out of the application band of interest [5].

The complexity of the construction of data converters in smaller feature size, combined with the demand for flexible architectures by modern computational systems is creating a vacuum for novel computing paradigms [6]. Neuromorphic computing suggests one such intriguing approach which can adaptively perform big amount of energy-efficient operations in parallel, such as pattern recognition [7]. Notably, analog-to-digital conversion can be seen as an example of simple pattern recognition, where the analog input can be classified into one of the $2^N$ different patterns for $N$ bits, and thus can be readily solved using artificial neural networks (ANNs). Furthermore, the calibration process of these networks can be viewed as modification of neural parameters based on the measured error calculated during learning [6].

In our previous work [2], we proposed a four-bit neural network (NN) ADC which can be trained in real-time. It implements artificial synapses using memristors [8] that are trained using an online stochastic gradient descent (SGD) algorithm until the ADC achieves optimal figure-of-merit (FOM). The training algorithm ensures power consumption optimization, mismatch self-calibration, and noise tolerance [9]. However, four-bit resolution is insufficient for practical applications [10]-[12], while direct scaling of this architecture is challenging due to the quadratic increase in number of synaptic weights (with exponentially large values), large area, high power consumption, longer training time, and limited sampling frequency.

This paper takes a different approach toward designing large-scale general-purpose neuromorphic ADCs. We propose a hybrid CMOS-memristor design with multiple trainable cores of four-bit NN ADC [2] and NN DAC [13] in a two-stage pipeline. This architecture takes advantage of light-weight low-power sub-ADC cores combined with high throughput and resolution achievable through the pipeline. Furthermore, each sub-ADC optimizes the effective number of bits (ENOB) and power dissipation during training for the chosen sampling frequency [2].

The remainder of this paper is organized as follows. In Section II, neuromorphic data converters and scaling challenges are discussed. In Section III, the proposed ADC architecture and the training framework are described. In Section IV, the pipeline ADC circuit is evaluated, and the paper is concluded in Section V.

## II. NEUROMORPHIC DATA CONVERTERS

### A. Neuromorphic ADC

The deterministic four-bit neural network ADC in [2] converts an analog input voltage ($V_{in}$) to a digital output code ($D_3D_2D_1D_0$) according to the following iterative expressions,

$$\begin{cases} D_3 = u(V_{in} - 8V_{ref}), \\ D_2 = u(V_{in} - 4V_{ref} - 8D_3), \\ D_1 = u(V_{in} - 2V_{ref} - 4D_2 - 8D_3), \\ D_0 = u(V_{in} - V_{ref} - 2D_1 - 4D_2 - 8D_3), \end{cases} \quad (1)$$

where Vref is the reference voltage equals to one full-scale voltage quantum (LSB), and $u(.)$ is the signum neural activation function (neuron) having either zero or full-scale voltage output. The neural network shown in Fig. 1(a) implements (1) in hardware using reconfigurable synaptic weights ($W_{i,j}$ – conductance between a pre-synaptic neuron with index $j$ and a post-synaptic neuron with index $i$) to address their non-deterministic distribution in real-time operation and post-silicon fabrication. As shown in Fig. 1(b), the synapses are realized using one NMOS, one PMOS and one memristor, with the transistor gates connected to a common enable input $e$ [14]. When $e = V_{DD}$ ($-V_{DD}$), the NMOS (PMOS) switches on and $u$ ($-u$) is passed to the output. When $e = 0$, both transistors are off and the output is zero. The neurons comprise of an inverting op-amp for integration and a latched comparator for decision making.

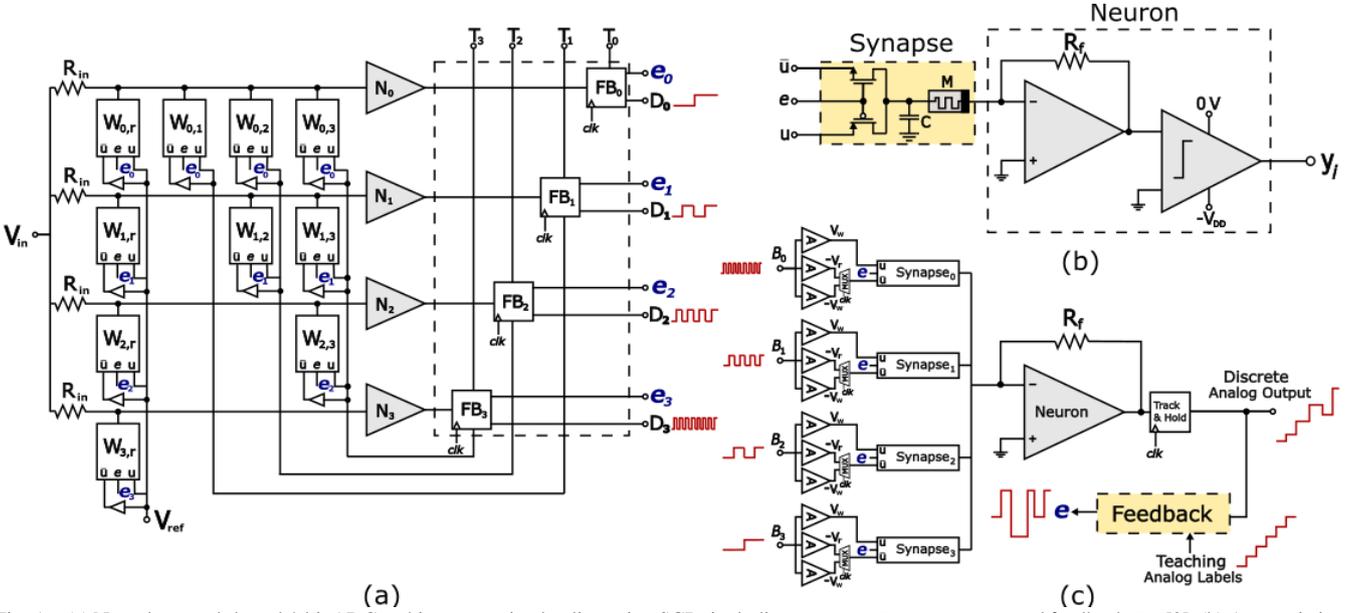

Fig. 1. (a) Neural network-based 4-bit ADC architecture trained online using SGD, including synapses $W_{i,j}$, neurons $N_i$, and feedback $FB_i$ [2], (b) A memristive synapse connected to an artificial neuron implemented as an inverting opAmp for integration and a comparator for decision making. (c) Neural network-based 4-bit DAC architecture, including synapse $W_i$, a neuron implemented as an opAmp, and a PWM-based feedback circuitry [13] for the time-varying SGD.

Synaptic weights are tuned to minimize the mean square error (MSE) by using the SGD learning rule,

$$\Delta W_{ij(j>i)}^{(k)} = -\eta \big(T_i^{(k)} - D_i^{(k)}\big) T_j^{(k)}, \quad (2)$$

where $\eta$ is the learning rate (a small positive constant), and in each iteration $k$, the output of the network $D_i^{(k)}$ is compared to the desired teaching label $T_i^{(k)}$ that corresponds to the input $V_{in}^{(k)}$. The training continues until the training error falls to $E_{threshold}$, a predefined constant that defines the learning accuracy. The FOM is optimized and the network is configured from a random initial state to the desired ADC.

### B. Neuromorphic DAC

The neural network DAC in [13] converts the four-bit digital input code ($V_3V_2V_1V_0$) to an analog output (A) as

$$A = \frac{1}{2^4} \sum_{i=0}^{3} 2^i V_i, \quad (3)$$

where binary weights ($2^i$) are implemented with reconfigurable synaptic weights $W_i$ and having similar realization as in Fig. 1(b). As shown in Fig. 1(c), the four synapses collectively integrate the input through the neuron (op-amp) to produce the output. This output is compared to the analog teaching labels in the pulse width modulation (PWM)-based feedback circuit, which regulates the value of the weights in real-time according to the time-varying gradient descent learning rule,

$$\Delta W_i^{(k)} = -\eta(t)\big(V_{out}^{(k)} - t^{(k)}\big) D_i^{(k)}, \quad (4)$$

where $\eta(t)$ is the time-varying learning rate, and $t^{(k)}$ is the analog teaching label. The feedback is disconnected after the training is complete ($E < E_{threshold}$).

### C. Scaling Challenges in Neuromorphic ADC

Increasing the scale of the neural network ADC, above four bits, is challenging. Table I highlights the effect of scaling on design and performance parameters of the ADC. The number of synapses in the network increases quadratically. Consequently, the area and power consumption rise significantly. Moreover, there is an exponential rise in the aspect ratio of synaptic weights, which is practically limited by the high-to-low resistive states ratio (HRS/LRS), number of resistive levels, endurance of the memristor (#trainings/day for ten years), and time and power consumption of the training phase [15] – ultimately limiting the practical achievable resolution to four-bits. Additionally, higher number of neurons require longer conversion-time which limits the maximal Nyquist sampling frequency.

TABLE I. SCALING CHALLENGES IN NEUROMORPHIC ADC

| Parameter | 4-bit | 8-bit | N-bit |
|---|---|---|---|
| # Neurons, feedbacks | 4 | 8 | $N$ |
| # Synapses | 10 | 36 | $N(N+1)/2$ |
| Total area (µm²) | 4850 | 9740 | $N(1.1N+1250)$ |
| Conversion rate (GSPS) | 1.66 | 0.74 | $1/(N \cdot t_p + (N-1)/BW)$ |
| Power (µW) | 100 | 650 | $P_{int} + P_{act} + P_{synapse}$ |
| FOM (fJ/conv) | 8.25 | 7.5 | $P/(2^{N-0.3} \cdot f_s)$ |
| HRS/LRS (memristor) | $2^4$ | $2^8$ | $2^{N-1+\log_2(Vdd/Vfs)}$ |
| # Levels (memristor) | 64 | 2048 | $N \cdot 2^N$ |
| # Training samples | 4000 | 6000 | $(2 - 2^{1-N/4}) \cdot 4000$ |
| Wearout (trainings/day) | 150 | 100 | $150/(2 - 2^{1-\frac{N}{4}})$ |

## III. NEUROMORPHIC PIPELINED ADC

### A. Neural Network Architecture

We propose using coarse-resolution neural network ADCs and DACs to build a fine-resolution pipelined network. An eight-bit two-stage pipelined network is shown in Fig. 2. In the first-stage sub-ADC, a synapse $W_{ij}$ is present between a pre-synaptic neuron with index $j$ and digital output $D_j$, and a post-synaptic neuron with index $i$, and output $D_i$. A neuron for each bit collectively integrates inputs from all synapses and produces an output by the signum neural activation function $u(.)$. The sub-ADC coarsely quantizes (MSBs) the sampled input $V_{in}$ to the digital code $D_7D_6D_5D_4$ as

$$\begin{cases} D_7 = u(V_{in} - 8V_{ref}), \\ D_6 = u(V_{in} - 4V_{ref} - W_{6,7}D_7), \\ D_5 = u(V_{in} - 2V_{ref} - W_{5,6}D_6 - W_{5,7}D_7), \\ D_4 = u(V_{in} - V_{ref} - W_{4,5}D_5 - W_{4,6}D_6 - W_{4,7}D_7). \end{cases} \quad (5)$$

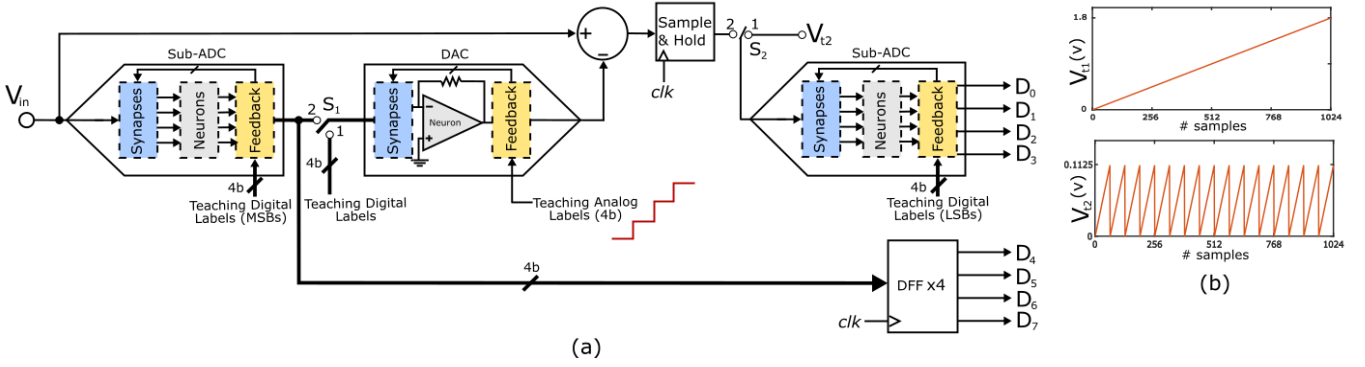

Fig. 2. (a) Proposed architecture of a two-stage pipelined ADC trained online using SGD. The first stage consists of four-bit single-layer neural network sub-ADC and DAC. The second stage consists of another four-bit neural network ADC. Both stages operate simultaneosuly to increase the conversion throughput and their intermediate results are temporarily stored in D-flipflop registers. (b) Training input $V_{t1}$ and $V_{t2}$ correspond to digital labels of two sub-ADCs.

The output of the sub-ADC is converted back to an analog signal $A$ by the DAC according to

$$A = \frac{1}{2^4} \sum_{i=4}^{7} W_i D_i , \qquad (6)$$

where $W_i$ are the synaptic weights. Next, this output is subtracted from the held input to produce a residue $Q$ as

$$Q = V_{in} - A . \qquad (7)$$

This residue is sent to the next stage of the pipeline, where it is first sampled and held. The second stage sub-ADC is designed similar to that of the first stage, except that the resistive weights of the input are modified from $R_{in} = R_f$ (feedback resistance of neuron) to $R_f/16$. This is made in order to scale the input from $V_{FS}/16$ to the full-scale voltage $V_{FS}$. The LSBs of the digital output are obtained from this stage as

$$\begin{cases} D_3 = u(16Q - 8V_{ref}), \\ D_2 = u(16Q - 4V_{ref} - W_{2,3}D_3), \\ D_1 = u(16Q - 2V_{ref} - W_{1,2}D_2 - W_{1,3}D_3), \\ D_0 = u(16Q - V_{ref} - W_{0,1}D_1 - W_{0,2}D_2 - W_{0,3}D_3). \end{cases} \qquad (8)$$

The sample-and-hold circuit enables concurrent operation of the two stages, achieving a high throughput rate, but introduces latency of two clock cycles. Thus D-flipflop registers are used to time-align the MSBs and the LSBs.

Conventional pipeline implementations generally use power-hungry flash sub-ADC cores and rely on redundancies and complex calibration techniques for high resolution [16]-[18]. On the other hand, trainable neural network ADC/DAC cores in this design have minimalistic design with mismatch self-calibration, noise tolerance, and power consumption optimization. This eliminates the need for an exclusive inter-stage gain unit and calibration mechanism, because the residue is amplified by the input resistive weight of the second sub-ADC. Although resistors are highly prone to manufacturing variations, they can be effectively used as the input weights because their mismatches will be calibrated for by other memristive weights in the second stage [13]. Furthermore, the training algorithm ensures that the quantization error remains within tolerable limits without using digital calibration techniques. This eliminates the area and power overheads of the calibration circuits, which overwhelm around 33% and 17% of the total area and power, respectively, in [18].

### B. Training Framework

The aim of the training is to configure the network from a random initial state (random synaptic weights) to an accurate eight-bit ADC. It is achieved by minimizing the mean-square-error (MSE) of each sub-ADC and the DAC by using specific teaching labels for desired quantization. During the training phase, switches $S_1$ and $S_2$ are in position 1.

TABLE II.    CIRCUIT PARAMETERS

| Parameter | Value | Parameter | Value |
|---|---|---|---|
| Power supply | | Feedback resistor | |
| $V_{DD}$ | 1.8 V | $R_f$ | 45 kΩ |
| NMOS | | PMOS | |
| W/L | 10 | W/L | 20 |
| $V_{TN}$ | 0.56 V | $V_{TP}$ | -0.57 V |
| Memristor | | | |
| $V_{on/off}$ | -0.3 V, 0.4 V | $R_{on/off}$ | 2 kΩ, 100 kΩ |
| $K_{on/off}$ | -4.8μm/s, 2.8μm/s | $\alpha_{on/off}$ | 3, 1 |
| Reading voltage and time | | Writing voltage and time | |
| $V_r$ | -0.1125 V | $V_w$ | ±0.5 V |
| $T_r$ | 5 μs | $T_w$ | 5 μs |
| Learning parameters | | Sub-ADC/DAC parameters | |
| $\eta_{ADC/DAC}$ | 1, 1 | $f_s$ | 0.1 MSPS |
| $E_{threshold\ ADC/DAC}$ | 4.5·10⁻², 9·10⁻³ | $V_{FS}$ | $V_{DD}$ |

The DAC is supplied with four-bit digital teaching labels corresponding to an analog ramp input, as shown in Fig. 2(a). We use the binary-weighted time-varying gradient descent rule in (4) to minimize the MSE between the estimated and desired label. Learning parameters are listed in Table II. The DAC is connected to the sub-ADC by switch $S_1$ when the error falls below $E_{threshold}$.

The accuracy requirements of each stage decrease through the pipeline and the first stage should be accurate to the overall resolution [18]. Moreover, the two-stages operate on different inputs for different quantization. Thus, their teaching dataset must be different to execute the online SGD algorithm as

$$\Delta W_{ij(j>i)}^{(k)} = -\eta_{ADC}(T_i^{(k)} - D_i^{(k)})T_j^{(k)}, 0 \leq i,j \leq 3, \quad (9)$$

$$\Delta W_{ij(j>i)}^{(k)} = -\eta_{ADC}(T_i^{(k)} - D_i^{(k)})T_j^{(k)}, 4 \leq i,j \leq 7. \quad (10)$$

Interestingly, (9) and (10) can be implemented using different teaching inputs, as shown in Fig. 2(b). Furthermore, the two stages can be trained independently and in parallel as their teaching datasets are supplied separately.

For the training dataset, an analog ramp signal is sampled at $4 \cdot 2^8$ (=1024). Four adjacent samples are given the same digital labels, providing an eight-bit training dataset, shown as $V_{t1}$ in Fig. 2(b). The more we train the ADCs with extra labels, the higher conversion accuracy we achieve. This is because of the nonlinear nature of the ADC task. The analog ramp input with the corresponding four MSBs is used to train the first stage sub-ADC. A sawtooth version of this input ($V_{t2}$ in Fig. 2(b)) with the remaining LSBs is used for the training of second stage. The switch $S_2$ is turned to position 2, when the overall mean-square-error falls below $E_{threshold}$.

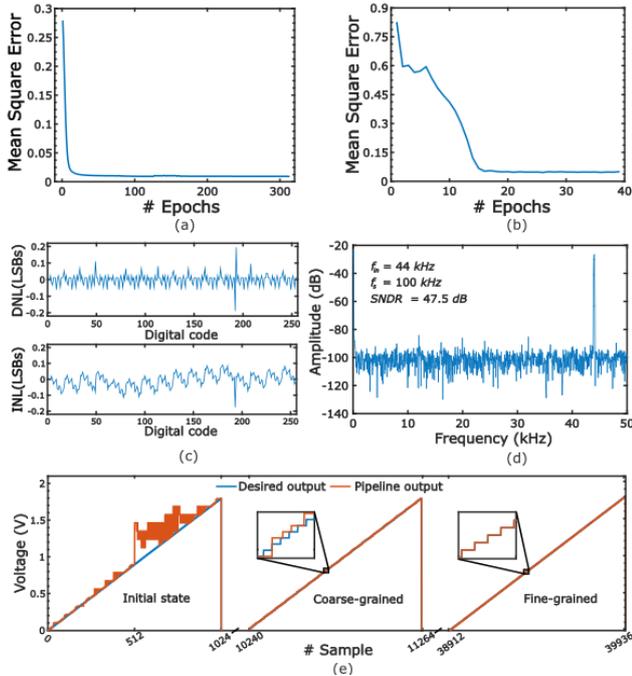

Fig. 3. (a) Pipeline ADC training evaluation. (a) Mean square error (MSE) of the first-stage DAC minimization during its training. (b) Total MSE of the two stages during training of sub-ADCs. (c) DNL and INL at the end of training. (d) 2048-point FFT for a 44 kHz sinusoidal input. (e) Comparison between the teaching dataset and the actual output of the ADC by connecting it to an ideal DAC, at three different timestamps during the training; an identical staircase is obtained when training is complete.

TABLE III. PERFORMANCE COMPARISON

| Parameter | NN ADC [2][a] | This work |
|---|---|---|
| # Bits | 8 | 8 |
| # Synapse | 36 | 24 |
| Memristor HRS/LRS | $2^8$ | $2^4$ |
| Max conversion rate $f_{max}$ (GSPS) | 0.74 | 1.66 |
| Power (μW) | 650 | 272 |
| FOM (fJ/conv) | 7.5 | 0.97[b] |
| Training time (ms) | 1060 | 400 |
| Wearout (trainings/day) | 25 | 55 |

[a] Based on scaling evaluation of 8b ADC with 4x training samples. [b] Extrapolated FOM at $f_{max}$.

TABLE IV. SCALABILITY EVALUATION

| # Bits | 12 |
|---|---|
| # Synapses | 38 |
| # Samples per epoch | $1 \cdot 2^{12}$ |
| Max |DNL| | 0.61 LSB |
| Max |INL| | 0.60 LSB |
| Training time (ms) | 2000 |

## IV. EVALUATION

Our proposed pipelined ADC is simulated and comprehensively evaluated in SPICE (Cadence Virtuoso) using a 180 nm CMOS process and memristors fitted by the VTEAM memristor model [19] to a Pt/HfO$_x$/Hf/TiN RRAM device [15]. The device has a HRS/LRS of 50. First, we evaluate the learning algorithm in terms of training error and learning time. Next, the circuit is statistically and dynamically evaluated, and finally, power consumption is analyzed. The circuit parameters are listed in Table II. To test the robustness of the design, we incorporate device non-idealities and noise, as listed in Table II in [13].

The basic deterministic functionality of the pipeline ADC is demonstrated during training by the online SGD algorithm. Fig. 3(a) shows the variation of the MSE of the first-stage DAC. After approximately 5,000 training samples (312 epochs), which equals 50 ms training time for a 0.1 MSPS conversion rate, the MSE error falls below $E_{threshold}$. Fig. 3(b) shows the total MSE of the two sub-ADCs. After approximately 40,000 training samples (39 epochs), which equals 400 ms training time, the total MSE falls below $E_{threshold}$. The analog output is converted through an ideal 8-bit DAC and probed at three different timestamps during training, as shown in Fig. 3(e). The output is identical to the input staircase after the training is completed.

Linearity plots (Fig. 3(c)), measured for 1.8 V ramp signal sampled by 18k points at 0.1 MSPS, show that dynamic nonlinearity (DNL) is within ± 0.20 LSB and integral nonlinearity (INL) is lower than ± 0.18 LSB. Fig. 3(d) shows the output spectrum at 0.1 MSPS sampling rate. The input is a 44kHz 1.8 $V_{pp}$ sine wave. The converter achieves 47.5 dB SNDR at the end of training. Next, we analyzed the power consumption of the network by considering neural integration power, neural activation power, and synapse power [2]. Remarkably, the total power consumption is optimized similar to [2] during training. The ADC consumes 272 μW of power, averaged over a full-scale ramp with $4 \cdot 2^8$ samples.

The reliability of the fitted memristive device has been studied in our previous work [2][13]. The endurance of the device is $8 \cdot 10^7$ cycles, which implies that the pipelined ADC could be reconfigured for ~55 times per day for ten years. The maximum conversion rate, $f_{max}$ = 1.66 GHz, is determined by the memristor cutoff frequency, the transit frequency of a 180 nm CMOS transistor, and the OpAmp slew rate [2], [13].

The proposed 8-bit pipelined architecture is compared to the scaled version of neural network ADC [2]. As shown in Table III, the pipelined ADC consumes less power, achieves high conversion rate, and better FOM with lesser HRS/LRS device ratio and number of resistive levels. To test the scalability of our architecture, we performed behavioral simulations in MATLAB. Our results for 12-bit design with ideal device parameters are summarized in Table IV. Furthermore, when the full-scale voltage is reduced to 0.9V and the sampling frequency is increased to 10 MSPS, the network converges to a new steady state to operate correctly under different specifications [2]. Currently, we are investigating logarithmic quantization training of the pipelined ADC for bio-medical applications [20], and leveraging other mixed-signal concepts such as delta-sigma modulation neurons for high-precision training [21].

## V. CONCLUSION

This paper proposed a novel pipelined neural network ADC architecture. This large-scale design was based on coarse-resolution neuromorphic ADC and DAC, modularly cascaded in a high-throughput pipeline and precisely trained online using SGD algorithm for multiple full-scale voltages, and sampling frequencies. The learning algorithm successfully tuned the neural network in non-ideal test conditions and configured the network as an accurate, fast, and low-power ADC. The hybrid CMOS-memristor design with 1.8 V full-scale voltage achieved 0.97 fJ/conv FOM at the maximum conversion rate. We believe that the proposed scalable architecture has promising results towards the realization of large-scale neuromorphic data converters for real-time adaptive applications.

ACKNOWLEDGEMENTS

This research was partially supported by the Israeli PBC Fellowship, by the Viterbi Graduate Fellowship and by European Research Council under the European Union's Horizon 2020 Research and Innovation Programme under Agreement 757259. The authors would like to thank Diana Ragozin for her help in simulations.

## REFERENCES

[1] Y. Chiu, B. Nikolic and P. R. Gray, "Scaling of Analog-to-Digital Converters into Ultra-Deep-Dubmicron CMOS," *Proceedings of the IEEE 2005 Custom Integrated Circuits Conference,* pp. 375-382, 2005.

[2] L. Danial, N. Wainstein, S. Kraus and S. Kvatinsky, "Breaking Through the Speed-Power-Accuracy Tradeoff in ADCs Using a Memristive Neuromorphic Architecture," *IEEE Transactions on Emerging Topics in Computational Intelligence*, Vol. 2, No. 5, pp. 396-409, Oct. 2018.

[3] S. H. Lewis, "Optimizing the Stage Resolution in Pipelined, Multistage, Analog-to-Digital Converters for Video-Rate Applications," *IEEE Transactions on Circuits and Systems II: Analog and Digital Signal Processing*, Vol. 39, No. 8, pp. 516-523, Aug. 1992.

[4] C. C. Lee and M. P. Flynn, "A SAR-Assisted Two-Stage Pipeline ADC," *IEEE Journal of Solid-State Circuits*, Vol. 46, No. 4, pp. 859-869, April 2011.

[5] N. N. Çikan and M. Aksoy, "Analog to Digital Converters Performance Evaluation Using Figure of Merits in Industrial Applications," *European Modelling Symposium*, pp. 205-209, 2016.

[6] E. O. Neftci, "Data and Power Efficient Intelligence with Neuromorphic Learning Machines," *iScience*, Vol. 5, pp. 52–68, 2018.

[7] A. Tankimanova and A. P. James, "Neural Network-Based Analog-to-Digital Converters," *Memristor and Memristive Neural Networks*, Apr. 2018.

[8] L. Chua, "Memristor-The missing circuit element,"*IEEE Transactions on Circuit Theory*, Vol. 18, No. 5, pp. 507-519, September 1971.

[9] L. Danial, and S. Kvatinsky, "Real-Time Trainable Data Converters for General Purpose Applications", *Proceedings of the IEEE/ACM International Symposium on Nanoscale Architectures*, July 2018.

[10] S. Ben Aziza, D. Dzahini and L. Gallin-Martel, "A High Speed High Resolution Readout with 14-bits Area Efficient SAR-ADC Adapted for New Generations of CMOS Image Sensors," *2015 11th Conference on Ph.D. Research in Microelectronics and Electronics (PRIME)*, pp. 89-92, 2015.

[11] A. Correia, P. Barquinha, J. Marques and J. Goe, "A High-Resolution Δ-Modulator ADC with Oversampling and Noise-Shaping for IoT," *2018 14th Conference on Ph.D. Research in Microelectronics and Electronics (PRIME)*, pp. 33-36, 2018.

[12] K. Garje, S. Kumar, A. Tripathi, G. Maruthi and M. Kumar, "A High CMRR, High Resolution Bio-ASIC for ECG Signals," *2016 20th International Symposium on VLSI Design and Test (VDAT)*, pp. 1-2, 2016.

[13] L. Danial, N. Wainstein, S. Kraus and S. Kvatinsky, "DIDACTIC: A Data-Intelligent Digital-to-Analog Converter with a Trainable Integrated Circuit using Memristors," *IEEE Journal on Emerging and Selected Topics in Circuits and Systems*, Vol. 8, No. 1, pp. 146-158, March 2018.

[14] D. Soudry, D. Di Castro, A. Gal, A. Kolodny and S. Kvatinsky, "Memristor-Based Multilayer Neural Networks With Online Gradient Descent Training," *IEEE Transactions on Neural Networks and Learning Systems*, Vol. 26, No. 10, pp. 2408-2421, Oct. 2015.

[15] J. Sandrini, B. Attarimashalkoubeh, E. Shahrabi, I. Krawczuk, and Y. Leblebici, "Effect of Metal Buffer Layer and Thermal Annealing on HfOx-based ReRAMs," *2016 IEEE International Conference on the Science of Electrical Engineering (ICSEE)*, pp. 1-5, Nov. 2016.

[16] D. Beck, D. Allstot, and D. Garrity, "An 8-bit, 1.8 V, 20 MSample/s Analog-to-Digital Converter Using Low Gain Opamps," *Proceedings of the 2003 International Symposium on Circuits and Systems, 2003*.

[17] P. Harpe, A. Baschirotto and K. A. A. Makinwa, *High-Performance AD and DA Converters, IC Design in Scaled Technologies, and Time-Domain Signal Processing*. Springer, 2014.

[18] C. Tseng, C. Lai and H. Chen, "A 6-Bit 1 GS/s Pipeline ADC Using Incomplete Settling With Background Sampling-Point Calibration," *IEEE Transactions on Circuits and Systems I: Regular Papers*, Vol. 61, No. 10, pp. 2805-2815, Oct. 2014.

[19] S. Kvatinsky, M. Ramadan, E. G. Friedman and A. Kolodny, "VTEAM: A General Model for Voltage-Controlled Memristors," *IEEE Transactions on Circuits and Systems II: Express Briefs*, Vol. 62, No. 8, pp. 786-790, Aug. 2015.

[20] L. Danial, K. Sharma, S. Dwivedi, and S. Kvatinsky, "Logarithmic Neural Network Data Converters using Memristors for Biomedical Applications", *Proceedings of the IEEE Biomedical Circuits and Systems (BioCAS)*, October 2019.

[21] L. Danial, S. Thomas, and S. Kvatinsky, "Delta-Sigma Modulation Neurons for High-Precision Training of Memristive Synapses in Deep Neural Networks", *IEEE International Symposium on Circuits and Systems (ISCAS),* May 2019.